\documentclass[sigconf]{acmart}

\usepackage{url}
\usepackage{code}
\usepackage{graphicx}
\usepackage{balance}
\usepackage{multirow}
\usepackage{multicol}
\usepackage{listings}
\usepackage{booktabs}
\usepackage{subcaption}
\usepackage{cleveref}
\usepackage{xspace}
\usepackage[most]{tcolorbox}

\hypersetup{draft}

\lstdefinestyle{PythonStyle}{
  language=Python,
  frame=single,
  backgroundcolor=\color{white},
  basicstyle=\footnotesize\ttfamily,
  keywordstyle=\color{blue}\bfseries,
  commentstyle=\color{darkgreen},
  stringstyle=\color{red},
  numbers=left,
  numberstyle=\tiny\color{gray},
  stepnumber=1,
  numbersep=5pt,
  showspaces=false,
  showstringspaces=false,
  showtabs=false,
  tabsize=2,
  captionpos=b,
  breaklines=true,
  breakatwhitespace=true,
  escapeinside={\%*}{*)},
  morekeywords={*,...}
}

\definecolor{mycolor}{RGB}{119,195,236}

\newtcolorbox{highlight}[1][ht]{
  colback=mycolor!10!white,
  colframe=mycolor!80!black,
  fonttitle=\bfseries,
  coltitle=mycolor!80!black,
  colbacktitle=mycolor!20!white,
  boxrule=1pt,
  arc=3pt,
  outer arc=3pt,
  boxsep=0pt,
  left=10pt,
  right=10pt,
  top=6pt,
  bottom=6pt,
  toptitle=2pt,
  bottomtitle=2pt,
  lefttitle=2pt,
  righttitle=2pt,
  titlerule=0pt,
  attach boxed title to top left={yshift=-\tcboxedtitleheight/2, xshift=2mm},
  boxed title style={arc=3pt, outer arc=3pt},
}

\begin{document}

%%
%% The "title" command has an optional parameter,
%% allowing the author to define a "short title" to be used in page headers.
\title{\toolname: Feedback-Driven, Agentic Test Suite Generation}

%%
%% The "author" command and its associated commands are used to define
%% the authors and their affiliations.
%% Of note is the shared affiliation of the first two authors, and the
%% "authornote" and "authornotemark" commands
%% used to denote shared contribution to the research.
%\author{Alex Groce}
%\affiliation{\institution{Northern Arizona University}\country{United States}}

%%
%% By default, the full list of authors will be used in the page
%% headers. Often, this list is too long, and will overlap
%% other information printed in the page headers. This command allows
%% the author to define a more concise list
%% of authors' names for this purpose.
\author{Kush Jain}
\affiliation{\institution{Carnegie Mellon University}
\country{United States}}

\author{Claire Le Goues}
\affiliation{\institution{Carnegie Mellon University}
\country{United States}}

\renewcommand{\shortauthors}{us folks}

%% Table shortcuts
\newcommand{\mr}[2]{\multirow{#1}{*}{#2}}
\newcommand{\mc}[3]{\multicolumn{#1}{#2}{#3}}

%% comments
\newcommand{\clg}[1]{\textcolor{blue}{#1}}
\newcommand{\kj}[1]{\textcolor{olive}{\textbf{\small[Kush: #1]}}}
\newcommand{\todo}[1]{\textcolor{red}{todo: #1}}

% To disable comments before submission, uncomment this:
% \newcommand{\clg}[1]{}
% \newcommand{\adg}[1]{}
% \newcommand{\kj}[1]{}
% \newcommand{\todo}[1]{}
% \newcommand{\uri}[1]{}

%% numbers
\newcommand{\toolname}{TestForge\xspace}

%%
%% The abstract is a short summary of the work to be presented in the
%% article.
\begin{abstract}

Automated test generation holds great promise for alleviating the burdens of manual test creation. However, existing search-based techniques compromise on test readability, while LLM-based approaches are prohibitively expensive in practice.
We present \toolname, an agentic unit testing framework designed to cost-effectively generate high-quality test suites for real-world code. Our key insight is to reframe LLM-based test generation as an iterative process.
\toolname thus begins with tests generated via zero-shot prompting, and then continuously refines those tests based on feedback from test executions and coverage reports. 
We evaluate \toolname on TestGenEval, a real world unit test generation benchmark sourced from 11 large scale open source repositories; we show that \toolname achieves a pass@1 rate of 84.3\%, 44.4\% line coverage and 33.8\% mutation score on average, outperforming prior classical approaches and a one-iteration LLM-based baseline.
\toolname produces more natural and understandable tests compared to state-of-the-art search-based techniques, and offers substantial cost savings over LLM-based techniques (at \$0.63 per file). 
%Our ablation study shows that both zero-shot test seeding and execution feedback iterations improve line-coverage by 9.6\%. 
Finally, we release a version of TestGenEval integrated with the OpenHands platform, a popular open-source framework featuring a diverse set of software engineering agents and agentic benchmarks, for future extension and development. 
% \clg{TestGenEval is used-before-definition here.  What is it? (I know, but your normal readers don't).}
% \clg{thought: should be careful with the phrasing of the readability claim, since we basically don't evaluate it (even though it's obvious.}

\end{abstract}

\setcopyright{none} % to remove the copyright notice
\settopmatter{printacmref=false} % to remove the ACM Reference Format
\renewcommand\footnotetextcopyrightpermission[1]{}

% \begin{CCSXML}
% <ccs2012>
% <concept>
% <concept_id>10011007.10010940.10010992.10010998.10011001</concept_id>
% <concept_desc>Software and its engineering~Dynamic analysis</concept_desc>
% <concept_significance>500</concept_significance>
% </concept>
% <concept>
% <concept_id>10011007.10011074.10011099.10011102.10011103</concept_id>
% <concept_desc>Software and its engineering~Software testing and debugging</concept_desc>
% <concept_significance>500</concept_significance>
% </concept>
% </ccs2012>
% \end{CCSXML}

% \ccsdesc[500]{Software and its engineering~Dynamic analysis}
% \ccsdesc[500]{Software and its engineering~Software testing and debugging}

% \keywords{test generation, large language models, mutation testing}

\maketitle

\section{Introduction}

\begin{figure*}[ht]
    \centering
    \includegraphics[width=\linewidth]{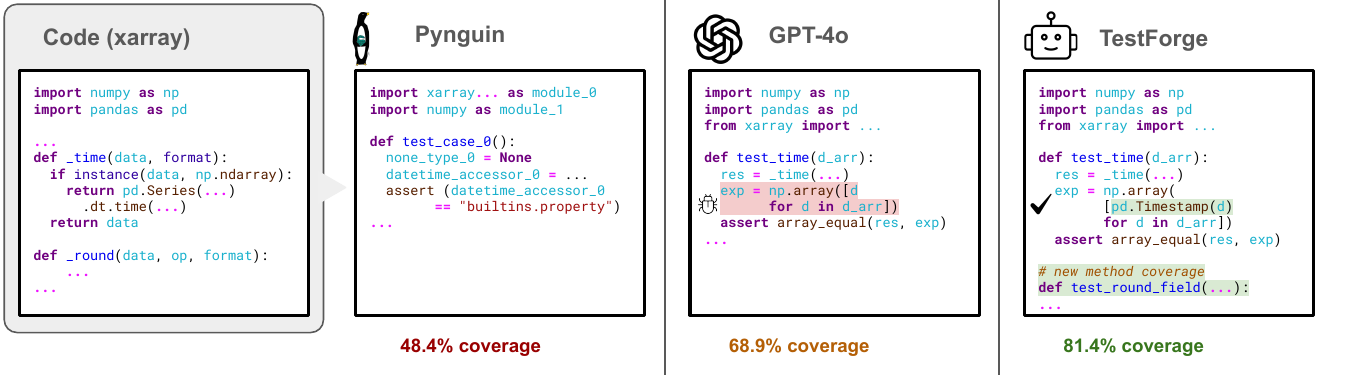}
    \vspace{-10pt}
    \caption{Example of tests generated by different approaches for timing and rounding functionality in pydata/xarray. GPT-4o generates a buggy test, which \toolname fixes, while also adding additional coverage improving tests.}
    \label{fig:motivating-example}
    \Description{Example of tests generated by different approaches for timing and rounding functionality in pydata/xarray. GPT-4o generates a buggy test, which \toolname fixes, while also adding additional coverage improving tests.}
\end{figure*}

Software testing plays a crucial role in software development; robust code bases typically include extensive unit and integration tests. However, creating effective tests is costly and time consuming~\cite{Beller1, Beller2}, causing developers to frequently ignore and neglect testing. As a result, there has been research into developing techniques for automated test generation~\cite{DinellaTOGA, FraserEvoSuite, BrandtEvoSuiteStudy, BaldoniSymbolicExecution, WatsonATLAS, VillmowContest}.

Classical test generation approaches like EvoSuite~\cite{FraserEvoSuite} and Pynguin~\cite{pynguin} focus primarily on improving test coverage using search-based techniques (e.g., genetic programming, random search). These methods are capable of producing relatively high-coverage test suites.  
However, they also often generate suites that are difficult to understand and maintain~\cite{panichella2020revisiting}. The resulting tests require substantial developer effort for debugging and comprehension. This, in turn, hinders overall adoption~\cite{BrandtEvoSuiteStudy}.

Meanwhile, recent advances in Large Language Models (LLMs) have demonstrated remarkable capabilities to generate human-readable code~\cite{CodeXGLUE, nijkamp2022codegen, fried2022incoder, bavarian2022efficient}. Tools like Copilot or Cursor have proven particularly effective for code generation, enhancing developer productivity~\cite{copilot,cursor}. 
That said, off-the-shelf LLMs perform relatively much less well on test generation tasks as compared to code generation for specific functionality; 
% \clg{There may be another reference besides cat-lm that can boost that argument; everyone I've ever talked about this with agrees that it's true. Honestly, probably your ICLR paper? Doesn't it show that everyone sucks at this?}
one reason may be the specific context required to write effective tests~\cite{catlm, jain2024testgenevalrealworldunit, ouedraogo2024largescaleindependentcomprehensivestudy}. Generating tests for large scale projects requires reasoning about the code under test, along with project and library dependencies used in the code under test. This limits prior approaches that only take in limited context such as the method or file under test~\cite{NieETAL22Teco, catlm, schäfer2023empiricalevaluationusinglarge}. Furthermore,  LLMs struggle with hallucination~\cite{liangUsabilityAIProgramming2024,jain2024testgenevalrealworldunit,liu2024exploring}; hallucinated test setup and outputs often cause generated tests to fail,  hindering adoption. 
That said, LLMs have demonstrated potential to improve classical techniques~\cite{codamosa} by boosting coverage in scenarios where genetic programming plateaus.
% \clg{...this paragraph did not originally articulate any limitations of LLM-based techniques, before I added the bit in the middle that currently cites catlm...I originally thought maybe a mention of cost, but actually, the prior work discussion on agent-based approaches gets there.  So instead, I think this paragraph should expand on the limitations of vanilla LLM-based approaches that makes agentic approaches better, since the next paragraph focuses on ``agents good''.  I think the bit I added about test generation specifically benefiting from added context (and anything else we can say in that argument) would be better to end on, because those are the things that an agentic approach are particularly good at.}

A particularly promising recent direction for overcoming some of these limitations of LLM-based test generation takes an \emph{agentic} approach. 
Agentic AI systems are characterized by the use of independent LLM-querying ``agents'' that have some degree of autonomy in choosing how to tackle a given task, and can interact with an environment providing access to additional tools like code search, static/dynamic analysis, or test execution. 
Such a system can autonomously explore code, edit tests, and self-reflect on its outputs. 
Recent techniques like CoverUp~\cite{pizzorno2025coverupcoverageguidedllmbasedtest} and 
HITS~\cite{wang2024hitshighcoveragellmbasedunit} have demonstrated the value of incorporating dynamic information (like coverage analysis) or 
more complex slicing techniques into an LLM-mediated test generation loop. 
These approaches are better at generating both passing and high coverage test suites compared to both classical and one-iteration LLM techniques, as they can iterate based on rich execution feedback.
% \clg{Is this true of gboth of them, or just CoverUp? The original description, it didn't sound like HITS did this. }
However, while previous agentic methods have demonstrated the benefits of iterative feedback, their applicability has been limited by high costs---over two dollars per large file---when applied to complex, real-world code bases~\cite{jain2024testgenevalrealworldunit}.

In this paper, we present \toolname, an agentic approach for  test generation that can cost-effectively generate unit tests for large files in complex code bases.
\toolname is predicated on three key insights. 
First, we frame test generation as an \emph{iterative} process.
\toolname begins by zero-shot prompting an underlying LLM model with the code under test. It then  progressively refines those tests over multiple iterations to target undercovered lines, or regions exhibiting low mutation scores.  Second, \toolname leverages detailed execution feedback---including compilation errors, runtime failures, and uncovered code segments---as an integral part of its agentic loop.
We provide the full set of lines missing coverage as input to agent rather than selecting a subset as in prior approaches. This enables the agent to plan out multiple test cases in one iteration, improving efficiency. This feedback-driven process both improves test quality, and contributes to a high empirical pass@1 rate in our evaluation.  
Finally, \toolname crucially operates at the file level, rather than the individual method level considered in previous LLM- or agentic-based approaches.  
This dramatically improves cost-efficiency---the average file in our benchmark includes 58 methods---without in practice compromising effectiveness. 

We evaluate \toolname's full test-suite-generation ability on the recently-released TestGenEval~\cite{jain2024testgenevalrealworldunit}, 
a benchmark for evaluating test generation over complex real-world systems.
For search-based techniques, we compare against Pynguin~\cite{pynguin} (pure genetic programming) and CodaMosa~\cite{codamosa} (genetic programming augmented with LLMs). We also compare against non-agentic LLM-based baselines CAT-LM~\cite{catlm}, and GPT-4o~\cite{openai2023gpt4}. Our experiments on the TestGenEval benchmark show that \toolname achieves a record pass@1 rate of 84.3\%, a line coverage of 44.4\%, and a mutation score of 33.8\% outperforming both LLM baselines and existing genetic programming approaches on the programs where the techniques apply. The differences in mutation score over baselines are even more pronounced than coverage differences; \toolname achieves a mutation score improvement of 15.4\% over our one-iteration baseline, improving the ability of generated test suites to catch synthetic bugs even when coverage is high.
% \clg{something about mutation score? I cut the redundant paragraph about real-world concerns, but the sentiment is still good and valid, and mutation score is a nice thing to include in punching up that argument.}
Because \toolname relies fundamentally on an LLM for code generation, the resulting tests are much more natural than those produced by classical search-based techniques. 
\toolname only costs \$0.63 to generate tests for a file in our dataset.  

Additionally, first, we integrate \toolname into the OpenHands\footnote{\url{https://github.com/All-Hands-AI/OpenHands}; Note that we omit a link to the pull request for anonymous review.}
open source platform for developing and evaluating autonomous agents; it empowers researchers to build modular, agentic while facilitating reproducible research in the field of AI-powered software testing.  
Second, as part of our evaluation, we integrate the TestGenEval benchmark into OpenHands as well, to better support reproducibility and future research.

In summary, our contributions are as follows: 
\begin{itemize} 
\item \toolname, a state-of-the-art and cost-effective test generation agentic system.  \toolname uses dynamic feedback and an interative approach to generate high-coverage and effective test suites for real-world code.  
\item Empirical results demonstrating \toolname's effectiveness compared to both classical and LLM-based baselines, in terms of pass@1 rate, coverage and mutation scores.
\item An ablation study of our design decision to start from the zero-shot generated test suite and an experiment measuring line coverage compared to the number of iterations of \toolname.
\item An integration of both \toolname and the TestGenEval benchmark with OpenHands, supporting reproducibility and extension, and evaluation of any new test generation system built with the OpenHands framework on the real-world benchmark.\footnote{Anonymized replication: https://anonymous.4open.science/
r/OpenHands-7E28/; we will link github repositories and pull requests in a final version of this paper.}
\end{itemize}

The rest of this paper is organized as follows: \Cref{sec:motex} provides an example showcasing key insights behind \toolname, \Cref{sec:approach} outlines the full design of \toolname, while \Cref{sec:experimentalsetup} and \Cref{sec:results} outline our evaluation setup and results respectively. 
% \clg{I know I'm the only person on Earth who likes roadmaps, and maybe we won't have one, but, just in case, how's about a placeholder here...}

\label{sec:approach}
\begin{figure*}
    \centering
\includegraphics[width=\linewidth]{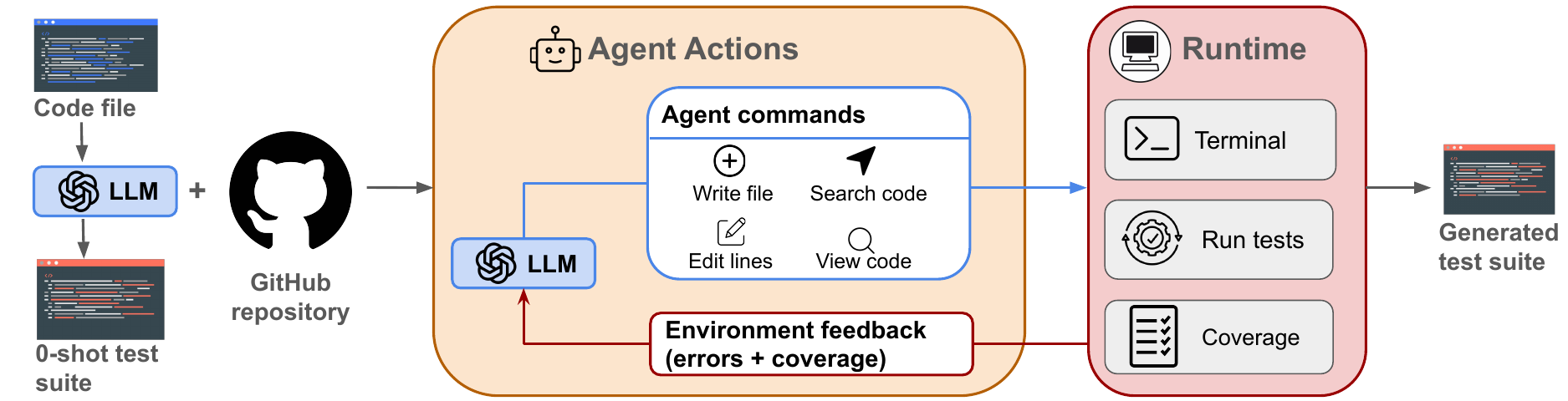}
    \caption{Overview of \toolname. We start by generating a zero-shot test suite and allowing our agent to interact with the repository with the generated test suite. We include the ability to search code, view code, write and edit files along with environment capabilities to run commands and tests. The output is a full test suite for the code file under test.}
    \label{fig:testgeneval-overview}
\end{figure*}

\section{Illustrative Example}
\label{sec:motex}

We begin by motivating and illustrating our approach with an example.  
The left-hand-side of \Cref{fig:motivating-example} shows a code snippet from \texttt{pydata/xarray},\footnote{https://github.com/pydata/xarray} a widely used Python package built on top of numpy that enables labeling and aggregation over multi-dimensional tensors. 
\toolname generates tests at the file-level; that is, it aims to unit test all 24 methods in a file.
The code under test in this file includes a number of utility accessor methods, such as rounding an array of objects, or converting a date object to a string time.

The second column of \Cref{fig:motivating-example} shows a test for this file produced by 
Pynguin~\cite{pynguin}, a search-based approach that uses a genetic programming heuristic to generate high-coverage unit tests for Python.  
Pynguin achieves 48.4\% line coverage and 0.9\% mutation score with a 10 minute compute budget on this file using four generated tests; the Figure shows the first such generated test.
In addition to low coverage and near zero mutation score for the generated suite, note, crucially, that this test is difficult to understand.  Methods and variables receive generic names, and the assertion is, plainly, difficult to follow and unrelated to the code under test (purpose is to assert the type of the instantiated variable).
We observe this pattern with other tests generated in our evaluation. Additionally, for this file, of four tests Pynguin generates, three fail due to issues with test inputs, such as trying to round a null variable or run greatest common denominator on a string.  
Pynguin cannot fix such tests, and instead simply (though correctly) marks them as expected to fail. 

Simply and directly zero-shot prompting 
GPT-4o~\cite{openai2023gpt4} with the code under test, and requesting unit tests for the file, results in 18 tests.
These tests achieve a coverage of 68.9\% and mutation score of 47.4\% on the file, outperforming Pynguin; the names and structure of this test better match our expectations for human-readable code. 
Despite improved coverage and mutation score, however, there are still methods in \texttt{pydata/xarray} that this one-iteration suite does not execute, like the \texttt{\_round} method.  
Additionally, hallucination and incorrectness are well-known problems with LLM-generated code, and GPT-4o generated tests frequently contain subtle bugs.  
Here, for example, the output \texttt{exp} checked in the assertion should be converted to a timestamp before the comparison. Furthermore, the issue is clear, with Python raising an \texttt{AttributeError} when running the test and even displaying the expected output with the time series data type as part of the message. The output from executing the test can therefore be easily leveraged to improve the overall test suite (one of the core insights behind \toolname).
% \clg{OK here's a thing: how can we tell this?  Do you need to manually look at it, or is it obvious due to a runtime exception?  I suspect it's the latter.  It's important to say this first because my first thought as the disinterested reader is "OK, do you need to manually check all tests?". Second, because I think the answer is "the execution gives feedback that we can use to improve the test" which is central to the method.}

Other LLM agent approaches such as HITS~\cite{wang2024hitshighcoveragellmbasedunit} and CoverUp~\cite{pizzorno2025coverupcoverageguidedllmbasedtest} use dependency analysis and program slicing as part of their agentic loop. For each method, these approaches generate tests and try to refine them individually to maximize coverage over all methods. While this works well for cases where source files are reasonably short and self-contained, it unfortunately does not scale well to projects with long contexts and complex dependencies. The full set of dependencies for a project such as \texttt{pydata/xarray} exceeds the 128k context window of most LLMs. This makes generating tests with both approaches very costly (over 3X the cost of \toolname), and thus not practical for these long context files; it is important to refine multiple tests at once rather than each test individually in order to save cost. 
% \clg{...I'm toying with suggesting a paragraph here that explains why we don't run one of the agentic approaches, for the comparison with the benefits of our approach.  I'm still wondering if there's a key novelty in the particular agentic setup that we have.  It's fine either way, but it's very much not clear right now, and it should be, from the intro (and maybe again mentioned here, where we discuss our insights).}

These observation motivate our approach in \toolname.  \toolname 
starts with the zero-shot solution as a template for iterative test refinement. These generated tests are generally both readable, and achieve moderate coverage on the code under test. This allows us to frame test improvement as an iterative process, with \toolname slowly improving the coverage and correctness of the generated tests. Second, \toolname leverages  execution feedback such as the missing lines in the coverage report and the test execution output. 
This allows the model to see and iteratively fix bugs in the initial generated tests. 
In this example, for example, \toolname can use the runtime error from the zero-shot test to fix the subtle bug and the missing lines in the coverage report for the \texttt{\_round} method to add a new test that covers this method.
In addition, the coverage report information provides targeted lines for the model to target, resulting in higher coverage of the refined test suites. The fourth column of \Cref{fig:motivating-example} shows the first test in the suite produced by \toolname for this example; it both fixes the bug in the GPT-4o tests, and the overall test suite covers previously untested methods, resulting in both a high coverage of 81.4\% and high mutation score of 54.4\% over the tested file. \toolname adds an additional 16 tests to the 18 GPT-4o-generated tests, with a total of 34 tests in the final test suite.

\section{\toolname}

In this Section, we provide an overview of \toolname, including the technical design details and the general agentic workflow. \Cref{sec:tooloverview} shows the 
design of \toolname and an illustrative example of the agentic workflow. \Cref{sec:toolactions} provides more technical details behind each 
tool call our agent can make, \Cref{sec:toolfeedback} provides an overview of all execution feedback we can get from the environment, and \Cref{sec:implementation} provides details about our implementation of \toolname and integration with OpenHands.

\subsection{Overview}
\label{sec:tooloverview}

\Cref{fig:testgeneval-overview} shows a full overview of \toolname. The input of \toolname is a code file to be tested; the output is a unit test suite for that file. Prior to the agentic loop, we prompt the LLM with the code under test, and ask it to generate an initial test suite, which we use as the starting test file (we show this is better than starting from scratch in our ablations (Section~\ref{sec:design-eval}). Following this, \toolname enters the agentic loop. \toolname can either perform an \emph{agent command} (actions listed in the orange box, labeled ``Agent Actions'') or \emph{interact with the environment} (actions listed  in the red box, labeled ``Runtime''). When \toolname is finished (signified by executing the bash command \texttt{exit} or 25 iterations have elapsed, we save the generated test suite.

We provide a list of available agent actions and their corresponding input and output formats as tools when calling OpenAI's API. Our actions are consistent with CodeAct~\cite{wang2024executablecodeactionselicit}, as we build off of the CodeAct agent.  If an action is malformed, we provide the agent with feedback corresponding to the command output (if the command fails altogether, we provide the tool parsing error). We follow a similar approach for environment interaction, by providing the agent feedback on whether the command ran successfully, along with the output of running the command. For example, if the agent runs the generated test suite, we provide feedback on whether all tests passed (and the command succeeded or not), along with the terminal output containing any test error messages. 

To encourage planning and reflection, we require \toolname to reflect on the output of any interaction with the environment. Self-reflection looks similar to chain-of-thought reasoning~\cite{cot}, where we show \toolname the command output and ask it to plan out subsequent steps systematically. For example, if the test suite generated by \toolname had three failing tests and failed to cover two lines, the agent is prompted to reflect, ideally producing a plan to fix the three failing tests and add a new test to cover the two missing lines.

\lstset{language=}

\begin{lstlisting}[caption=TestForge prompt template, label=lst:prompt]
Your goal is to improve the test suite at {test_file} to achieve **broad-coverage** of the code below.

...
IMPORTANT REQUIREMENTS:
1. Check coverage after each iteration
2. No external help or resources use only the snippet below.
...
Below is the **complete code snippet** to test:
{code_src}
...
Output the final test suite (20+ tests) for {test_file} in a single code block
\end{lstlisting}

\Cref{lst:prompt} shows an abridged version of our agent prompt. We define the high level goal (generating a high coverage test suite) and provide the agent with a set of requirements for what a good test suite looks like and limitations of the environment (for example, no external dependencies). We prompt \toolname to generate greater than 20 tests, as this generally corresponds with higher mutation score and coverage.

\begin{lstlisting}[caption=TestForge bash tool call JSON definition, label=lst:toolcalltemplate]
 {
  "type": "function",
  "function": {
    "name": "execute_bash",
    "description": "Execute a bash command in the terminal...",
    "parameters": {
      "type": "object",
      "properties": {
         "command": {
           "type": "string",
            "description": "The bash command to execute..."
         },
         ...
      },
      "required": [
        "command"
      ]
    }
  }
}
\end{lstlisting}

\begin{lstlisting}[caption=TestForge bash tool call LLM output, label=lst:toolcallexample]
{
    "id": "call_h3lCE3GtZaKOke9hqPCdv0Wy",
    "type": "function",
    "function": {
        "name": "execute_bash",
        "arguments": "{"command": "..."}"
    }
}
\end{lstlisting}

\Cref{lst:toolcalltemplate} shows how we define both our agent actions and interactions with the environment in LiteLLM\footnote{https://github.com/BerriAI/litellm} as tools. This JSON definition is passed as input to the OpenAI API, which supports tool calling functionality. \Cref{lst:toolcallexample} shows an example output of calling the \texttt{execute\_bash} tool defined in \Cref{lst:toolcalltemplate}; the LLM supplies the required arguments and the name of the tool.

% \clg{I'm pretty sure you're gonna hate this suggestion, but I really want to know more about the prompt.  I don't want the full prompt, but a sketch or outline would be nice.  I just feel like, if I wanted to reimplement what you did from this description (true in section 3 in general), I would be wholly unable to.  I know, that's why we provide replication packages, but I should be able to at least get the GIST of what the paper did, otherwise, why have an approach section at all? Can we try a small figure showing a template or sketch or something? Or a flow showing an example? Oh, that might work: snippets of the steps TestForge takes to generate the tests in the illustrative example, maybe?  Which tools it uses? Pieces of the prompts? I don't know if it would work best to interleave the illustration with the description of the approach or do it all at once in a new subsection at the end...my initial instinct is the former, but I could be wrong.}

\subsection{Agent Actions}
\label{sec:toolactions}

We define the list of agent actions needed to navigate and edit complex code bases. We use the tool calling 
functionality of LiteLLM, which asks the LLM to generate a JSON object that represents a tool call. These tool calls are defined as a Python function, which performs the described functionality (for example viewing or editing a file). When the model outputs a well-formed JSON object representing the tool call, LiteLLM invokes the correct Python funtion. This is consistent with OpenHands's CodeAct agent~\cite{wang2024executablecodeactionselicit}, which achieves state-of-the-art performance on SWEBench~\cite{jimenez2024swebench}. 

% \clg{As what, like, python programs? Calls to command line things?}
% \clg{I uncommented this todo because I still have the question. 
%  Maybe the question is secretly asking, wtf is LiteLLM doing.  But the fact that someone else uses some other API to do this to achieve good performance on a different task doesn't answer my question.}
% CLG says: much better
 
\subsubsection{Navigate repository} An important function for code agents is navigating \textit{complex} code bases. We provide \toolname with the ability to search for files containing a particular prefix or suffix. We also instrument this with the ability to search across the entire repository or in a specific directory of the repository.

\subsubsection{Write file} We provide \toolname the ability to create any file or overwrite any existing file. This is also in line with prior work~\cite{yang2024sweagentagentcomputerinterfacesenable, wang2024executablecodeactionselicit}. In practice, \toolname primarily writes Python test files. There is no limit on how long the generated file can be. This enables \toolname to generate tests in cases where the zero-shot test suite has many errors by overwriting the buggy test file. 

\subsubsection{Edit file} We provide editing functionality to \toolname for any file (also in line with prior work~\cite{yang2024sweagentagentcomputerinterfacesenable,wang2024executablecodeactionselicit}). \toolname primarily edits Python test files. The command takes in both original text and replacement text.  The original text should occur once in the file. If the original text is not found in the file or occurs multiple times, the command fails, and the agent is prompted for another replacement. If the command succeeds, the changed lines are outputted to the agent, enabling iteration if the changes were not what was intended. We also provide functionality to insert lines into an existing file, enabling agents to add or insert new tests into an existing test file.

\subsubsection{View file} Our view file tool works in tandem with our search functionality. An agent can view a range of 400 lines in a file (with the option to specify the range). Often \toolname wants to target a specific range, for example, a specific set of lines not covered by the existing test suite, which can be done by specifying the appropriate range. We choose 400 lines to enable \toolname to understand other files in the repository, while not overwhelming the model context with file content.  This is consistent with both SWEAgent~\cite{yang2024sweagentagentcomputerinterfacesenable} and CodeAct~\cite{wang2024executablecodeactionselicit}.

\subsection{Environment Feedback}
\label{sec:toolfeedback}

In addition to agentic actions, the system also provides feedback from the environment to \toolname. We define environment feedback in LiteLLM in the same way as agent actions. Here, we outline sources of feedback for \toolname when generating tests.

\subsubsection{Test execution feedback} One important source of feedback is the output from test execution. We provide the agent with the command to execute the tests. Once tests are executed, we provide the full output from the execution of the tests, including any syntax or assertion errors in the generated test suite (assertion errors often contain expected output, making it easier for the agent to update the tests). Syntax errors can also be fixed by using the edit functionality. The output for the developer-provided tests fits in the 128k context of most LLMs, enabling us to include most test outputs in our prompt context.

\subsubsection{Code coverage feedback} In addition to test execution feedback, we also provide the coverage report feedback. We provide the agent with the command to run the code coverage report. Our coverage report feedback includes the file, set of covered lines and a set of lines that are missing coverage. We use the coverage library in Python to measure line coverage and generate the coverage report. The agent can then use this information to view uncovered lines in the file under test and add tests that target these lines. 

\subsubsection{Bash command feedback} We also provide \toolname the ability to execute arbitrary \texttt{bash} commands and receive execution feedback from running them. We run all commands in a docker image with the dependencies for \toolname and the individual project to mitigate any associated safety risks. The output from this tool call includes whether the command succeeded or failed, and all terminal output (stdout and stderr).

\subsection{Implementation}
\label{sec:implementation}

\toolname is implemented in approximately 4.3k LOC of Python; our anonymous replication package is available at: \url{https://anonymous.4open.science/r/OpenHands-7E28/}. To replicate our results, see the README at \texttt{evaluation/benchmarks/testgeneval/README.md}.  

We integrate \toolname into OpenHands, a framework intended to support modular building and extension of software engineering agents, while facilitating reproducible AI agent research. We use GPT-4o as our model of choice, due to high performance on TestGenEval~\cite{jain2024testgenevalrealworldunit} and easy API access. However, because we use LiteLLM, it is easy to switch which model we use. We also adapt TestGenEval to work with any agent integrated with the OpenHands framework~\cite{wang2024executablecodeactionselicit}. Our replication package includes the prompt for \toolname, code to run \toolname and the full agentic version of TestGenEval. We hope these contributions will enable others in the community to easily extend \toolname and evaluate future test generation agents.

\section{Experimental Setup}
\label{sec:experimentalsetup}
\begin{figure}[ht]
  \centering
  \centering
  \begin{subfigure}[b]{0.9\linewidth}
    \centering
    \includegraphics[width=\linewidth]{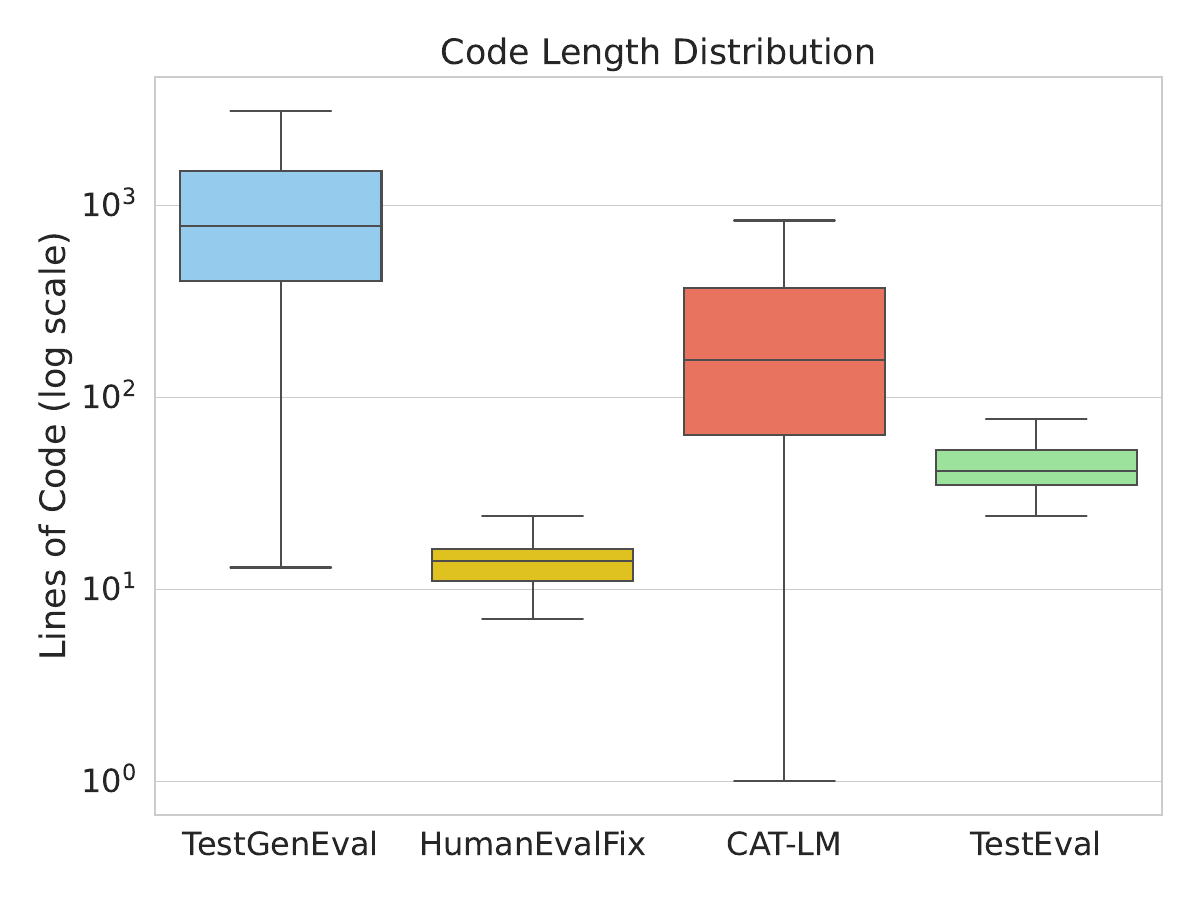}
    \vspace{-10pt}
    \caption{TestGenEval code lengths}
    \label{fig:swebenchcodelens}
  \end{subfigure}
  \vfill
  \begin{subfigure}[b]{0.9\linewidth}
    \centering
    \includegraphics[width=\linewidth]{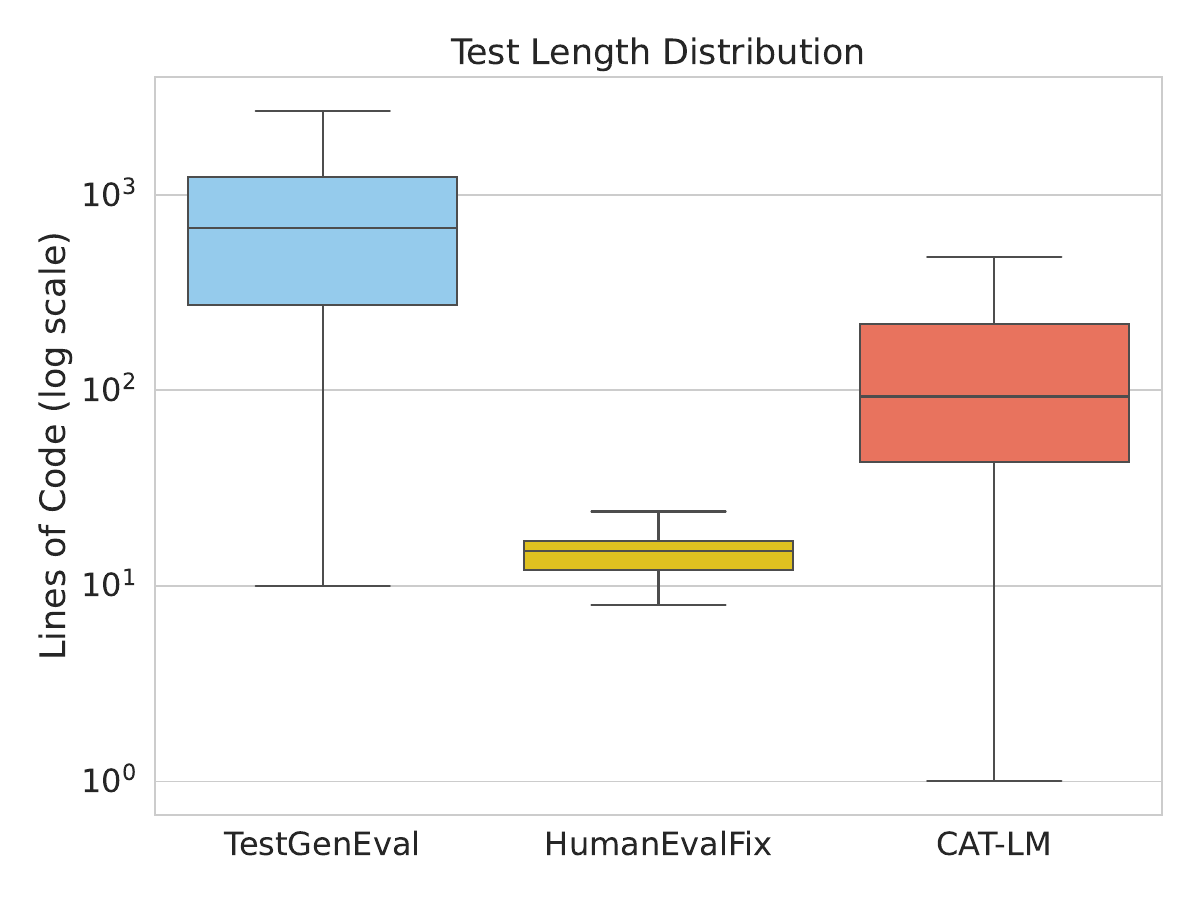}
    \vspace{-10pt}
    \caption{TestGenEval test lengths}
    \label{fig:swebenchtestlens}
  \end{subfigure}
  \vspace{-5pt}
  \caption{Code and test lengths across TestGenEval, HumanEvalFix, CAT-LM, TestEval. Code and test files in TestGenEval are significantly longer than other benchmarks (even with the log scale). TestEval is not included in the test lengths plot, as it does not contain ``gold'' tests.}
  \label{fig:testgenevallengths}
\end{figure}

We compare \toolname with Pynguin~\cite{pynguin} and CodaMosa~\cite{codamosa}, current state-of-the-art unit test generation tools and GPT-4o 0-shot prompting on the TestGenEval benchmark~\cite{jain2024testgenevalrealworldunit}. We ask the following research questions:

\noindent\textbf{RQ1: Runtime Performance: How well does \toolname perform at generating high coverage test suites?} We measure the pass@1, coverage and mutation score on a large scale benchmark of complex GitHub projects. We compare \toolname against Pynguin, CodaMosa, and zero-shot prompting, measuring the ability of each approach to generate full test suites.

\noindent\textbf{RQ2: Lexical Performance: How similar are test suites generated by \toolname to actual developer test suites?} We also measure lexical similarity of generated tests to developer written tests to understand whether tests generated by \toolname and other LLM approaches are more ``human like'' than search-based genetic programming approaches.

\noindent\textbf{RQ3: Design Decisions: How does performance vary with number of iterations? Does performance improve by starting with the zero-shot solution?} We measure the effect of varying the maximum number of iterations of agentic feedback on coverage. We also examine the insight that starting with the zero-shot solution is better than starting from scratch by performing an ablation of coverage and mutation score between these two approaches.

\noindent\textbf{RQ4: Behavior: Which actions does \toolname take most frequently? How does the readability and maintainability of tests generated by \toolname compare against other baselines?} We perform both a quantitative analysis of actions taken by \toolname and a small case study of tests generated by \toolname and each of the baselines. We discuss the readability and maintainability of test suites for each approach.

\subsection{Dataset}

We evaluated all baselines and \toolname on TestGenEval~\cite{jain2024testgenevalrealworldunit}, a benchmark of code test file pairs, sourced from 11 large-scale open source repositories (3,523-78,287 stars). \Cref{fig:testgenevallengths} shows the distribution of code and test lengths across TestGenEval, HumanEvalFix, CAT-LM, and TestEval. Code and test files are much longer in TestGenEval than other benchmarks: code files are on average 1157 LOC and test files are on average 943 LOC, while for CAT-LM, the benchmark with the second-longest files, code files are on average 344 LOC and test files are on average 211 LOC. This simulates a much more realistic experience of writing tests for complex repositories.
% \clg{Problem with this sentence: the ``much more realistic experience of...'' construction implies a ``much more THAN'' something.  But there's no comparison here.  Can you talk about general ballpark numbers for the other datasets you mention below? Better yet, can you provide side-by-side histograms? I understand if the answer is 'no' to the second question. But there needs to be SOME contextualization of these numbers.}

We choose TestGenEval over existing benchmarks such as CAT-LM~\cite{catlm}, TestEval~\cite{wang2024testevalbenchmarkinglargelanguage} and HumanEvalFix~\cite{chen2021evaluating}, as these existing benchmarks primarily target smaller code and test files, simple problems such as LeetCode and small programming problems. As a result, even zero-shot approaches saturate these benchmarks (coverage values greater than 85\% for GPT-4o)~\cite{wang2024testevalbenchmarkinglargelanguage,shi2024codecorrectnessclosingmile}. TestGenEval provides a benchmark that existing models perform poorly on, where we can measure the impact of execution feedback.

\subsection{Baselines}

We compare \toolname against multiple classical and LLM baselines. We run \toolname for 25 iterations, and classical baselines for 600 seconds (the average time per data point of \toolname is 447 seconds).  For classical baselines, we choose Pynguin~\cite{pynguin} and CodaMosa~\cite{codamosa}. Pynguin uses genetic programming to search the input space and maximize code coverage over the code under test, monitoring the output of the code under test with each generated input. These input and output pairs are then converted into test cases. CodaMosa~\cite{codamosa} extends this by using LLMs when Pynguin hits a coverage plateau (the same coverage for 25 iterations). We upgrade the model used in CodaMosa from Codex~\cite{CodeXGLUE} to GPT-4o~\cite{openai2024gpt4technicalreport} (Codex is no longer available in the OpenAI API).

% \clg{OK sorry I won't fix this for you, but I want to get through the notes I have: the following paragraph doesn't make sense yet because you haven't discussed dataset.  Either swap these two subsections, or move and integrate this paragraph into the next subsection (I think I have a minor preference for the latter but the feeling isn't strong).}
Both Pynguin and CodaMosa are only listed as compatible with Python 3.10. Of the 1210 programs in TestGenEval, only 45 use Python 3.10. We compare against both baselines on this subset of data. Furthermore, both baselines do not successfully generate tests for all 45 programs; CodaMosa only generates tests for 28 programs, and Pynguin generates tests for a different 27 programs. Errors include segmentation faults with both tools or issues with dependency analysis for complex projects (we raised an issue but the fix is not simple).\footnote{Issue elided for double blind purposes} For programs where either baseline fails to generate a test, we mark pass@1, coverage and mutation score as 0. 

In addition to these classical baselines, we compare against two LLM baselines: GPT-4o 0-shot~\cite{openai2023gpt4} and CAT-LM~\cite{catlm}. GPT-4o 0-shot provides a baseline for agentic improvement, as we start with 0-shot tests with \toolname. We use the \texttt{gpt-4o-2024-08-06} version of GPT-4o. CAT-LM is a state-of-the-art LLM for test generation, trained on an aligned data set of code and test files. By pretraining with this aligned set, CAT-LM is able to outperform much larger models with larger pretraining budgets. For both LLM baselines we follow the original TestGenEval paper~\cite{jain2024testgenevalrealworldunit}, using the same prompt and temperature of 0.2. Since both LLM baselines work on the full TestGenEval benchmark, we measure performance both on the entire 1210 programs and the subset of 45 programs that are compatible with Pynguin and CodaMosa.

Other LLM baselines exist as well, including MuTAP~\cite{dakhel2023effectivetestgenerationusing}, HITS~\cite{wang2024hitshighcoveragellmbasedunit} and CoverUp~\cite{pizzorno2025coverupcoverageguidedllmbasedtest}, however we were not able to compare against them. Unfortunately, MuTAP is exclusively integrated with HumanEvalFix, and cannot take arbitrary context; we therefore cannot trivially evaluate it on TestGenEval. HITS targets Java 17 and is not is not compatible with Python. Both HITS and CoverUp struggle with long context methods; these approaches iteratively refine coverage for a specific focal method, which does not scale to large files such as those in TestGenEval with a large number of methods. CoverUp costs approximately \$2 per data point in TestGenEval, due to operating at the method level. Even with \toolname that costs \$0.63 cents per file our experiments cost \$762 for all of TestGenEval; running CoverUp would cost \$2400, which is out of our price range.

\subsection{Metrics}

We measure test adequacy with both runtime and lexical metrics. Runtime metrics approximate the quality of generated test suite, while lexical metrics measure similarity between generated test suites and developer-written test suites. We report the cost of \toolname in USD to ensure that our approach is economically viable.
% \clg{This isn't grammatical; much of this subsection reads like it was written in a bit of hurry. ;-) Something funny is going on in the first sentence of the Pass@1 paragraph too.  Do a careful pass for me.  While you're at it, add context where necessary.  i.e., What is coverage measuring? (hint: is a (mediocre) proxy for test suite quality).  Segues into mutation score nicely.  When you get to lexical metrics, again remind me what they're for/why they're reasonable.}

\subsubsection{Runtime Metrics}

% \clg{I think the reason I don't feel like we prominently report cost is that it's never mentioned in metrics.  So, something about that here would be nice.}
\noindent\textbf{Pass@1} measures whether the generated test suite has \textit{any} test that passes for the code under test. We can add the resulting test suite to the existing code base by removing all failing tests. High pass@1 indicates that generated tests can be added to a test suite, but does not provide any guarantee of the quality of generated tests.

% \clg{...I kinda want to know the proportion of generated tests that fail, or maybe the percentage of tests per suite that pass on average.  Is this an interesting number? Is it something we can find?}

\noindent\textbf{Coverage} measures the proportion of code lines executed by passing tests in generated test suite. We omit failing tests from the coverage computation; these tests would require developer modification before being added to an existing test suite. Coverage serves as a weak proxy for test quality; high coverage indicates a test suite that executes most of the code under test, but does not guarantee the written tests have meaningful assertions.

\noindent\textbf{Mutation score} measures the percentage of synthetic bugs injected detected by the test suite (the test suite should pass on the original code and fail on the buggy code). To compute mutation score, we introduce synthetic bugs into the code under test and measure if the tests can detect these bugs. We use cosmic ray\footnote{https://github.com/sixty-north/cosmic-ray} as our mutation testing tool and use the standard set of operators. We also set a one hour timeout in line with TestGenEval (which only has a 1.06\% error in mutation score results).  Mutation score provides a more robust measure of test suite quality than other metrics~\cite{Ooutt1996SubsumptionOC,renMutantsSubsitute}. High mutation score indicates a test suite is capable of catching future bugs that may be introduced into the code under test, and thus is relatively robust. However, mutation score is costly to compute, as we have to run the entire test suite for each bug.

\subsubsection{Lexical Metrics}

\noindent\textbf{CodeBLEU}~\cite{ren2020codebleumethodautomaticevaluation} serves as a proxy of similarity between code snippets. Based on BLEU score, CodeBLEU considers n-gram match between both code pieces. It also incorporates code specific similarity measures such as the BLEU score of reserved keywords, AST match, and data flow match. 

\noindent\textbf{ROUGE}~\cite{lin-2004-rouge} measures the longest overlapping subsequence of tokens between the generated test and gold test, using F1 score. ROUGE has been used in prior testing work~\cite{catlm,NieETAL22Teco} as a metric of code similarity; high ROUGE indicates substantial overlap between developer written tests and generated tests.

\noindent\textbf{Edit Similarity} is the single character similarity between the generated test suite and gold standard developer-provided test suite for a file under test. Specifically it is 1 - Levenstein edit distance -- number of single character edits needed to transform the generated test to the gold test, normalized by the total number of characters. High edit similarity indicates similar tests to developer written tests, while low edit similarity indicates different tests. 

\section{Results and Analysis}
\label{sec:results}

% \clg{It's entirely possible that this thought is addressed somewhere, but as it occurred to me in reading the illustrative example, I'm leaving a note to check whether this should be discussed here: we say that Pynguin generates many tests with test setup issues.  Can we compare in some way/pull those results out in some detail?}

We report results for each research question and discuss their implications compared to other test generation approaches.

\subsection{RQ1: Runtime Performance}

\begin{table}
\centering
\begin{tabular}{@{}lrrr@{}}
\toprule
\textbf{Model}           & \textbf{Pass@1} & \textbf{Coverage} & \textbf{Mutation Score} \\ \midrule
% \textbf{CAT-LM} & 0.0\% & 0.0\% & 0.0\% \\ 
\multicolumn{4}{c}{\textbf{Full TestGenEval (1210 Programs)}} \\\midrule
\textbf{GPT-4o (0-shot)} & 64.0\% & 34.8\% & 18.4\% \\ 
\textbf{\toolname} & \textbf{84.3}\% & \textbf{44.4\%} & \textbf{33.8\%} \\ \midrule
\multicolumn{4}{c}{\textbf{Python 3.10 TestGenEval (45 Programs)}} \\\midrule
\textbf{GPT-4o (0-shot)} & 97.8\% & 54.0\% & 27.3\% \\ 
\textbf{Pynguin} & 60.0\% & 24.0\% & 4.2\% \\ 
\textbf{CodaMosa} & 62.2\% & 30.2\% & 2.7\% \\ 
\textbf{\toolname} & \textbf{100.0\%} & \textbf{60.0\%} & \textbf{31.6\%} \\ 

\bottomrule
\end{tabular}
\caption{Runtime results on the full TestGenEval dataset and Python 3.10 subset. \toolname achieves higher pass@1, coverage and mutation score than all baselines. All differences between baselines and \toolname are statistically significant (p < 0.05) other than GPT-4o 0-shot (p = 0.07) for the Python 3.10 subset.}
\vspace{-20pt}
\label{tab:baseline_comparison}
\end{table}

% \begin{table}
% \centering
% \begin{tabular}{@{}lrrr@{}}
% \toprule
% \textbf{Model}           & \textbf{Pass@1} & \textbf{Coverage} & \textbf{Mutation Score} \\ \midrule
% % \textbf{CAT-LM} & 0.0\% & 0.0\% & 0.0\% \\ 
% \textbf{GPT-4o (0-shot)} & 97.8\% & 54.0\% & 28.1\% \\ 
% \textbf{Pynguin} & 60.0\% & 24.0\% & 4.2\% \\ 
% \textbf{CodaMosa} & 62.2\% & 30.2\% & 2.7\% \\ 
% \textbf{\toolname} & \textbf{100.0\%} & \textbf{60.0\%} & \textbf{31.6\%} \\ 
% \bottomrule
% \end{tabular}
% \caption{\clg{What I would love is a way to combine tables 1-2 and 3-4 into single figures, or tables with subtables, or something.  I dunno.  Feels busy with 4 tables and 4 captions.} TestGenEval runtime results on 45 Python 3.10 programs. \toolname achieves higher coverage and mutation score than all baseline approaches. All differences between baselines and \toolname are statistically significant (p < 0.05) other than GPT-4o 0-shot (p = 0.07).}
% \label{tab:baseline_comparison_subset}
% \end{table}

\begin{table}
\centering
\begin{tabular}{lrrr}
\toprule
\textbf{Model} & \textbf{CodeBLEU} & \textbf{ROUGE} & \textbf{Edit Sim} \\
\midrule
\multicolumn{4}{c}{\textbf{Full TestGenEval (1210 Programs)}} \\\midrule
\textbf{CAT-LM} & 29.0 & 6.8 & 25.2 \\ 
\textbf{GPT-4o (0-shot)} & 31.8 & 22.9 & 25.9 \\ 
\textbf{\toolname} & \textbf{32.1} & \textbf{23.0} & \textbf{27.0} \\ \midrule
\multicolumn{4}{c}{\textbf{Python 3.10 TestGenEval (45 Programs)}} \\\midrule
\textbf{CAT-LM} & 29.6 & 4.8 & 21.8 \\ 
\textbf{GPT-4o (0-shot)} & \textbf{35.0} & 28.6 & 25.8 \\ 
\textbf{Pynguin} & 18.2 & 9.7 & 14.2 \\ 
\textbf{CodaMosa} & 11.3 & 10.0 & 17.5 \\ 
\textbf{\toolname} & 33.3 & \textbf{29.3} & \textbf{26.9} \\ 
\bottomrule
\end{tabular}
\caption{TestGenEval lexical results for all 1210 programs in the dataset. All LLM-based approaches achieve similar scores on all lexical metrics (besides the low ROUGE score of CAT-LM), but genetic programming approaches achieve far lower lexical performance.}

\label{tab:lexical_baseline_comparison}
\end{table}

\begin{figure}[ht]
  \centering
  \centering
  \begin{subfigure}[b]{0.9\linewidth}
    \centering
    \includegraphics[width=\linewidth]{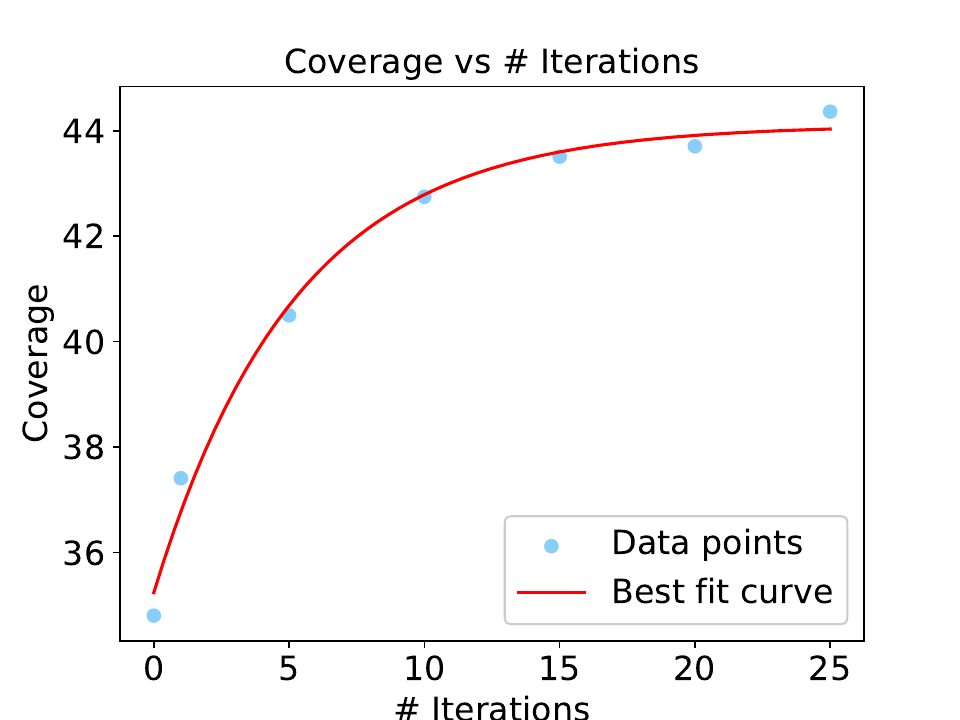}
    \caption{Coverage with \toolname iterations}
    \label{fig:coverageatk}
  \end{subfigure}
  \vfill
  \begin{subfigure}[b]{0.9\linewidth}
    \centering
    \includegraphics[width=\linewidth]{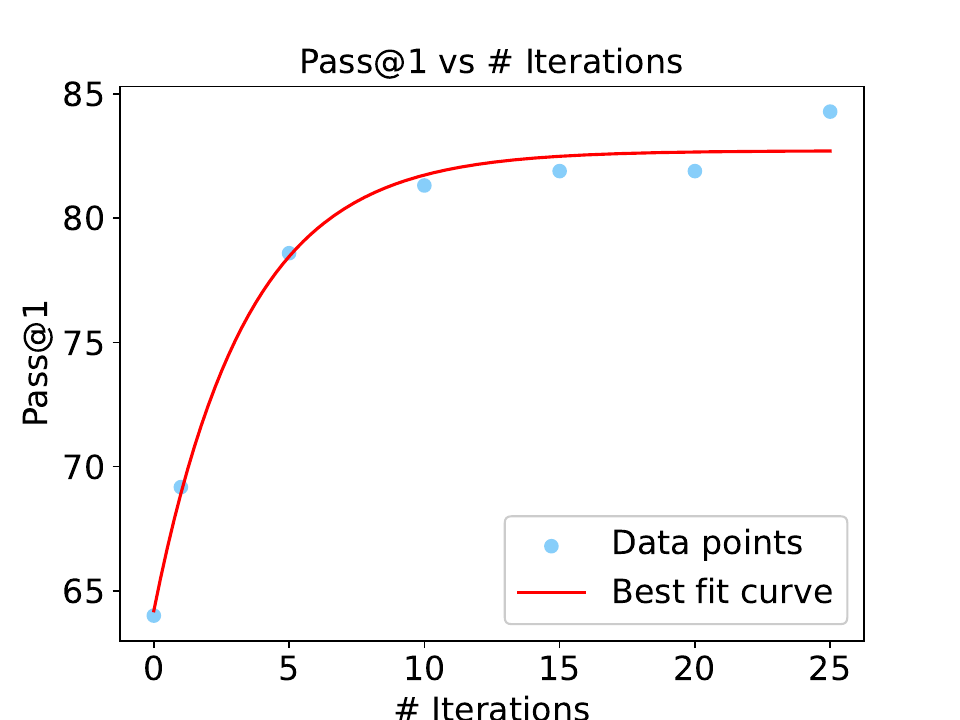}
    \caption{Pass@1 with \toolname iterations}
    \label{fig:passatk}
  \end{subfigure}
  \vspace{-5pt}
  \caption{Coverage and pass@1 in comparison to number of iterations. Both metrics have diminishing returns as k, increases, indicating k=25 is a good value.}
  \vspace{-15pt}
  \label{fig:metrick}
\end{figure}

\Cref{tab:baseline_comparison} shows the pass@1, coverage and mutation score of the test suites generated by GPT-4o and \toolname on all 1210 programs in TestGenEval. Pass@1 is relatively high for \toolname, with \toolname generating tests for 84.3\% of all programs. 0-shot prompting does significantly worse, only generating passing tests for 64.0\% of all programs. Evaluations of other previous techniques on benchmarks like HumanEvalFix or TestEval tend to produce higher coverage results; we note the relative complexity of the code in TestGenEval compared to these other benchmarks.  That said, \toolname outperforms 0-shot prompting by 9.6\%. This indicates that even for complex test suites, models still benefit from both execution feedback and coverage reports detailing missing lines. The generally low coverage can be attributed to the long code files present in TestGenEval, which are frequently 10,000+ tokens. Mutation score varies more between both approaches, with \toolname outperforming 0-shot prompting by 15.0\%. This indicates that using an agentic approach not only results in more code paths being executed in the code under test, but also in higher quality tests for the covered code. CAT-LM performs very poorly with 0\% pass@1, coverage and mutation score. CAT-LM is a very small LLM (approximately 3B parameters), therefore does not have the same test generation capabilities of larger models. All observed differences in results are statistically significant (p < 0.05).

We also report runtime metrics in comparison against CodaMosa and Pynguin on the 45 data point subset that uses Python 3.10 (a requirement for both CodaMosa and Pynguin). For cases where Pynguin and CodaMosa do not generate tests, we mark both the coverage and mutation score of the respective approach as 0\%. CAT-LM performs very poorly on this subset as well with 0\% pass@1, coverage and mutation score. For this subset of TestGenEval, pass@1 tends to be relatively high with both 0-shot and \toolname achieving nearly 100\% pass@1. 

Pynguin and CodaMosa struggle more, not generating tests for all cases, largely due to issues with analyzing dependencies for these complex programs. Coverage also varies significantly between approaches; Pynguin and CodaMosa struggle to generate high coverage test suites for these complex programs (even with the suggested 600 second budget for test generation in CodaMosa~\cite{codamosa}). Mutation score is even lower for baseline approaches, with genetic programming approaches primarily optimizing for high code coverage rather than mutation score. As a result, \toolname generates higher coverage test suites both genetic programming approaches by greater than 10\% and 6\% when compared to 0-shot. The difference for mutation score is even greater with a greater than 25\% difference between \toolname and both search-based baselines. 

Even excluding cases where Pynguin and CodaMosa fail to generate tests, \toolname outperforms both approaches with a coverage difference of 20.9\% and 12.2\% respectively and a mutation score difference of 23.2\% and 30.8\% respectively. All observed performance differences in results are statistically significant other than between \toolname and 0-shot, where the p-value is 0.07.

Finally, \toolname generates test suites in its default configuration at an average cost of \$0.63 per file under test, which we argue is reasonable efficient.  We evaluate the impact of running fewer iterations (providing potential cost savings) in Section~\ref{sec:design-eval}.

\subsection{RQ2: Lexical Performance}

We report lexical performance as a measure of how similar generated test suites are to developer written test suites. \Cref{tab:lexical_baseline_comparison} shows the CodeBLEU, ROUGE and edit-similarity of all LLM approaches. Lexical scores are generally low for TestGenEval; the ``gold standard'' developer-written test files are relatively long, comparatively, and existing models struggle to generate long test suites that are comprehensive. However, the generated tests are relatively natural and similar to developer-written tests, especially when compared to search-based test generation approaches. LLMs are pretrained on human code, whereas search-based approaches have no notion of ``naturalness''.

We also report lexical metrics for both LLM and search-based approaches on the Python 3.10 subset of TestGenEval. LLM approaches have CodeBLEU, ROUGE and edit-similarity of approximately 30\%, while search based approaches perform worse on all metrics by greater than 9\%. CodaMosa and Pynguin also perform poorly across all metrics; in general tests generated by these approaches use poor variable names and inputs that are more complex than typically present in human-written tests. \toolname and 0-shot slightly outperform CAT-LM, but the differences are relatively minor, with little to no difference between 0-shot and \toolname. 

\subsection{RQ3: Design Decisions}
\label{sec:design-eval}

\begin{table}
\centering
\begin{tabular}{@{}lrr@{}}
\toprule
\textbf{Model}           & \textbf{Pass@1} & \textbf{Coverage} \\ \midrule
% \textbf{CAT-LM} & 0.0\% & 0.0\% & 0.0\% \\ 
\textbf{\toolname (no seeding)} & 79.0\% & 42.1\% \\ 
\textbf{\toolname} & \textbf{84.3}\% & \textbf{44.4\%} \\ 
\bottomrule
\end{tabular}
\caption{TestGenEval runtime results on all 1210 programs in the dataset, with and without 0-shot seeding. Removing the 0-shot seeding results in a 2.3\% drop in coverage.}
\label{tab:ablation_testgeneval}
\vspace{-30pt}
\end{table}

We next evaluate the impact of design decisions involved in \toolname. Specifically, we look at the impact of (1) the number of agent iterations on test suite quality, and (2) providing the zero-shot test file as a starting point for \toolname.  We evaluate these in terms of coverage and pass@1 scores for the resulting test suites. 

\subsubsection{Number of iterations} We measure the performance as a function of the number of iterations of \toolname. The average cost per iteration is only four cents, making our approach relatively low-cost even as we scale up the number of iterations. More iterations provide more opportunities for \toolname to iterate on execution feedback and target lines that lack coverage.

\Cref{fig:metrick} shows both coverage and pass@1 as we increase the number of iterations. Coverage increases significantly in the first 10 iterations (almost 10\%), but after these 10 iterations the improvement in coverage is much more incremental. In a cost-constrained setting, one could use \toolname with only 10 iterations, halving the cost with only minimal performance loss. Pass@1 follows a similar trend, with significant gains in the first five iterations and only incremental improvement afterwards. The curve for pass@1 is steeper than coverage because the criteria is less fine-grained (if only one generated test passes in the test suite, pass@1 is 1).

\subsubsection{Removing the zero-shot starting test file} Another key insight behind \toolname is starting from the 0-shot generated test suite rather than from scratch. Intuitively, this allows the model to further improve coverage in cases where 0-shot prompting produces a reasonable test suite, while overwriting the existing test file in cases where the 0-shot generated test suite is far from correct. To make the comparison fair, we provide \toolname without seeding with an additional iteration.

\Cref{tab:ablation_testgeneval} shows the coverage and pass@1 of \toolname with and without zero-shot seeding. Both pass@1 and coverage go down without the seeding of the 0-shot file, with a 5.3\% drop in pass@1 and a 2.3\% drop in code coverage. 0-shot solutions are often not far from correct and can easily be iterated on to generate high quality test suites.

\subsection{RQ4: Behavior}
\label{sec:qual}

We perform both a quantitative analysis of actions taken by \toolname (\Cref{sec:quantaction}) and a qualitative analysis of tests generated by \toolname and other baselines (\Cref{sec:qualcasestudy}). 
\lstset{language=Python}

\begin{figure}
    \centering
\includegraphics[width=\linewidth]{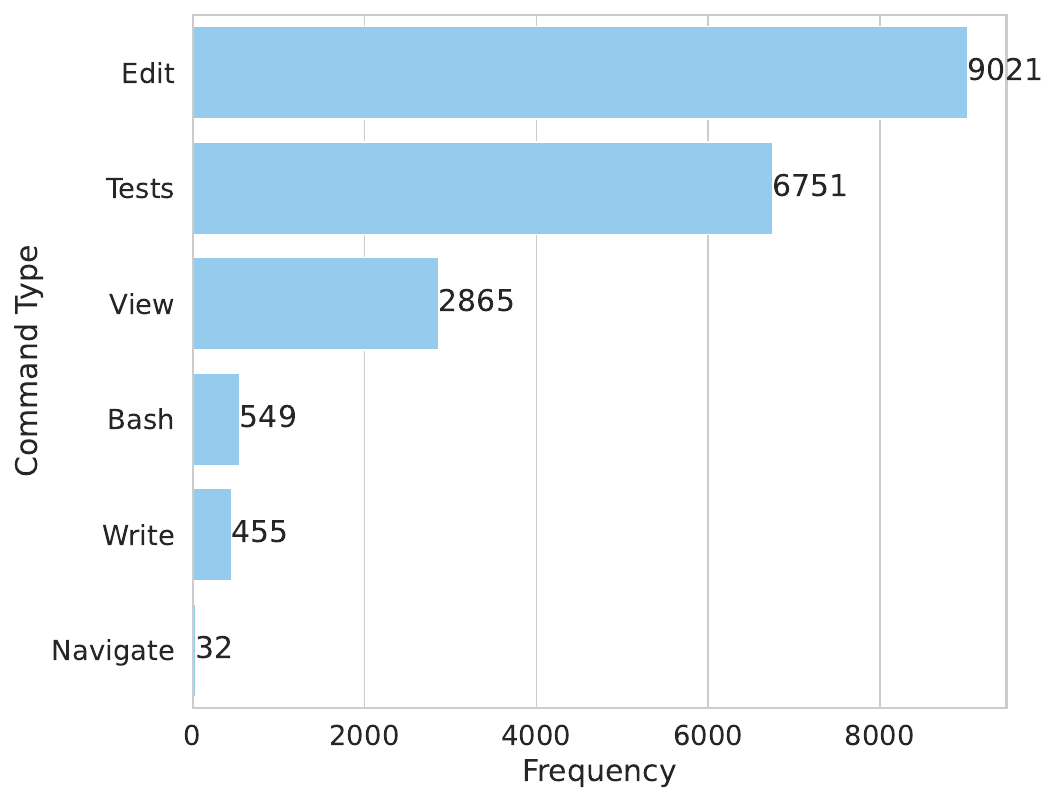}
\vspace{-20pt}
    \caption{Frequency of \toolname commands taken while generating tests for all 1210 programs in TestGenEval. The most common actions are editing and executing the generated test suite, indicating an iterative approach to test suite refinement.}
    \label{fig:testgeneval-actions}
\end{figure}

\subsubsection{Quantitative Action Analysis}
\label{sec:quantaction}

% \clg{As in the caption figure: I don't understand the dataset this is drawn from so I don't know how to interpret the results.  Is this over all 1210 programs?}
\Cref{fig:testgeneval-actions} shows the full set of actions taken by \toolname while generating tests for all 1210 programs in TestGenEval. ``View'' corresponds to viewing files, while edit, navigate and write correspond to editing the contents of a file, navigating the repository and writing a new file. ``Tests'' correspond to executing the generated test suite and obtaining both test and coverage feedback. Finally, ``Bash'' refers to executing arbitrary bash commands. The most frequent actions taken by \toolname are editing and executing the test suite (and also obtaining coverage information). We hypothesize the large number of edits and test execution corresponds to \toolname iterating on execution feedback (similar to what we see in \Cref{fig:motivating-example}). Navigation is much less frequently used, suggesting that the complex dependency analysis used by prior work~\cite{pizzorno2025coverupcoverageguidedllmbasedtest,wang2024hitshighcoveragellmbasedunit} might not be as important as code editing. In cases, where \toolname used navigation functionality, it was in the first few iterations, and dominated by code edits and test execution later. The most frequent bash scripts we observed were invocations to the Python interpreter (e.g. \texttt{python -c ...}), indicating \toolname uses our bash functionality to understand the execution behavior of the code under test. 
% \clg{Kinda interesting that navigation is so infrequent.  I feel like a bit more speculation could happen here --- maybe that expensive dependency analysis used by those other tools isn't as important as editing?}
% \clg{Maybe this question is impossible to answer so at some point you can stop going down this rabbit hole, but maybe...navigation happens in early iterations and so is dominated by editing actions over subsequent iterations?}
% \clg{What sorts of bash scripts?}

\subsubsection{Qualitative Readability Analysis}
\label{sec:qualcasestudy}
\begin{figure}[!th]
\centering
\begin{minipage}{0.45\textwidth}
\begin{lstlisting}[caption=Pynguin test, label=lst:cat-pynguinfirst]
@pytest.mark.xfail(strict=True)
def test_case_0():
    var_0 = module_0.__dir__()
    module_1.convert_label_indexer(var_0, var_0, tolerance=var_0)
\end{lstlisting}

\begin{lstlisting}[caption=CodaMosa test, label=lst:codamosafirst]
def test_case_2():
    tuple_0 = ()
    vectorized_indexer_0 = module_0.VectorizedIndexer(tuple_0)
    assert vectorized_indexer_0 is not None
    assert module_0.dask_array_type == ()
    assert len(module_0.integer_types) == 2
\end{lstlisting}

\begin{lstlisting}[caption=GPT-4o test, label=lst:gpt4first]
def test_convert_label_indexer():
    index = pd.Index([1, 2, 3])
    assert indexing.convert_label_indexer(index, 2) == (1, None)
    assert indexing.convert_label_indexer(index, slice(1, 3)) == (slice(1, 3), None)
\end{lstlisting}
\begin{lstlisting}[caption=TestForge (repaired) test, label=lst:testforgefirst]
def test_convert_label_indexer():
    index = pd.Index([1, 2, 3])
    assert indexing.convert_label_indexer(index, 2) == (1, None)
    assert indexing.convert_label_indexer(index, slice(1, 3)) == (slice(0, 3), None)
\end{lstlisting}
\end{minipage}
\vspace{-10pt}

\caption{Tests generated for the \texttt{label\_indexer} method. Pynguin and CodaMosa generate tests that are very hard to maintain (poor variable names, unintuitive values). GPT-4o generates a good test, with a subtle bug that \toolname fixes.}
\label{fig:toolcompletions}
\end{figure}

% pydata__xarray-3114-16452, pydata__xarray-4629-16492, pydata__xarray-7203-16577

Following prior work~\cite{catlm}, we randomly select three programs in TestGenEval where all baselines successfully generate tests with coverage. We highlight an example of tests generated by \toolname, and all baselines, discussing the trade-offs from a perspective of readability and maintainability. The other two examples can be found in Supplementary Materials.

\noindent\textbf{Pynguin:} Pynguin generates a test that covers the method under test. However, the test contains many attributes that hinder maintainability. The test is named poorly (\texttt{test\_case\_0}) and variables are named unintuitively. The test is also marked as expected fail, when it isn't clear what error or exception is being tested. Even the imports are named poorly (with the module being called \texttt{module\_0}). This is an inherent limitation of search based approaches; since they are not trained to mimic developer tests, the generated tests tend to look very different (also seen with low lexical metric scores).

\noindent\textbf{CodaMosa:} CodaMosa suffers from many of the same issues as Pynguin, because it is built off Pynguin. CodaMosa prompts LLMs to escape coverage plateaus in Pyguin's genetic search based approach. However, despite prompting LLMs, CodaMosa converts the variable names and inputs to the format that Pynguin uses. In this case, CodaMosa fails to generate a test directly covering the method under test. The other test shown suffers from the same problems as Pynguin (poor variable names, poor test name, importing the module as \texttt{module\_0}).

\noindent\textbf{GPT-4o 0-shot:} Unlike other baselines, GPT-4o generates a well-structured test. The GPT-4o generated test uses appropriate test and variable names. The inputs and expected outputs are also well formatted and easy to understand, with module imports using the correct names (pd for pandas). However, unfortunately the test does not pass due to a subtle bug in the expected output (should be \texttt{slice(0, 3)} instead of \texttt{slice(1, 3)}).

\noindent\textbf{\toolname:} \toolname generates a passing test that is both readable and easy to maintain. GPT-4o provides a good starting point for test generation, with a relatively good test structure and readable assert statements. By fixing the bug in the zero-shot generation \toolname can improve overall test suite coverage, while also producing an easy to maintain test that can be added to a developer test suite.

\section{Limitations}

% \clg{Note for later, so I don't forget; this might not belong here, not sure yet, I'm still in intro: we may want to tease out in discussion the differences in method-level vs. file-level coverage vs. cost.}
% \clg{Note for later: would be very convenient if it happened to be super impossible to generate file-level test suites using the previous techniques.  If so, we should say so, and why.  Probably not here, but rather where we describe baselines.  Just putting it here because the note needs to go somewhere.}

We outline potential limitations with \toolname and discuss their impact on our reported results.

\noindent\textbf{Data contamination:} A limitation of \toolname is that GPT-4o might have seen the TestGenEval test set at pretraining time. However, currently this does not seem to be a major issue, as even state-of-the-art agents still achieve relatively low performance on TestGenEval. Furthermore, we measure performance improvements of \toolname in comparison to the base model, making data leakage less of a concern, as we are concerned about relative performance differences rather than absolute performance values. The larger and newer models also have lower data contamination rates due to the large number of tokens present at the pretraining time~\cite{ramos2024largelanguagemodelsmemorizing}.

\noindent\textbf{Generalization of findings:} Another limitation is that our findings might not generalize to all repositories. We specifically target Python repositories with \toolname, and benchmark on TestGenEval, which is adapted from the SWEBench dataset. While the code and test files in TestGenEval are sourced from complex GitHub projects, the repositories in TestGenEval are widely popular. This might make them easier to test than other domain-specific or company specific repositories.

\noindent\textbf{Oracle problem:} One other limitation of our approach (and most automatic test generation approaches) is the oracle problem. One assumption behind \toolname is that the code under test is correct. However this might not be the case, as generated tests may fail on the code under test, while exposing bugs in the code under test. Despite this, our approach is still useful in catching regressions or bugs introduced in future versions of the code under test.

\section{Related Work}
\label{sec:related}

\noindent{\textbf{Classical Test Generation:}} Classical test generation techniques employ both black box and white box techniques to find ``breaking'' inputs for the code under test. Random/fuzzing techniques such as Randoop~\cite{PachecoRandoop}, aflplusplus~\cite{FioraldiAflplusplus} and honggfuzz maximize coverage by preserving inputs that improve coverage. Property testing tools such as Korat~\cite{BoyapatiKorat}, QuickCheck~\cite{koenquickcheck} and Hypothesis~\cite{MacIver2019Hypothesis} enable the systematic checking of user-defined properties. PeX~\cite{TillmannPex} and Eclipser~\cite{ChoiEclipser} use symbolic execution to mathmatically reason over a large input space and produce crashing input. The issue with both fuzzing and symbolic execution is they rely on the program crashing to find failing inputs, limiting the extend they can test programs. Search based test generation approaches~\cite{FraserEvoSuite, pynguin, codamosa} solve this problem by assuming the code under test is correct. These approaches maximize coverage by exploring test inputs that improve coverage and creating a large test suite of these inputs. However, search-based approaches struggle to generate ``natural'' tests, suffering from both stylistic and readability problems~\cite{BrandtEvoSuiteStudy, DakaEvoSuiteUserStudy, RobertsonEvoSuite}.

\noindent{\textbf{Neural Test Generation:}} More recently, neural test generation methods leverage language models to generate readable and natural tests. ConTest~\cite{VillmowContest} employs a generic transformer model that utilizes the tree representation of code to generate assert statements. Building on this, several approaches~\cite{WatsonATLAS,White2020ReAssertDL,TufanoAthenaTest,DinellaTOGA,hossain2024togllcorrectstrongtest} leverage transformer architectures for assertion generation. Their results demonstrate that LLM generated assertions are more natural and preferred by developers compared to existing tools such as EvoSuite. TeCo~\cite{NieETAL22Teco} further broadens the scope of the completion of the test by generating test statements one at a time. CAT-LM~\cite{catlm} leverages a LLM pretrained on aligned code and test filepairs to generate high quality unit tests. While these neural approaches address many readability challenges inherent in classical test generation methods, they primarily focus on generating individual tests, offering significantly fewer time-saving benefits.

\noindent{\textbf{Large Language Models of Code:}} Large language models (LLMs) can perform well across many software engineering tasks~\cite{brown2020languagegpt3, nijkamp2022codegen,nijkamp2023codegen2,grattafiori2024llama3herdmodels}. TestPilot~\cite{schäfer2023empiricalevaluationusinglarge} uses GitHub's Copilot to generate unit tests. GPT-4o is a state of the art LLM, with top performance on 
 many benchmarks~\cite{openai2024gpt4technicalreport,jain2024testgenevalrealworldunit,CodeXGLUE}. Empirical studies have evaluated LLMs efficacy in code and test generation, as well finding that performance varies heavily between benchmark and code base~\cite{yang2024evaluationlargelanguagemodels,schäfer2023empiricalevaluationusinglarge}

\noindent\textbf{Test Generation Agents:} Recently, there has been extensive work on developing software engineering and software testing agents. SWEAgent~\cite{yang2024sweagentagentcomputerinterfacesenable} leverages execution feedback to better solve software engineering pull requests. ChatUniTest~\cite{chen2024chatunitestframeworkllmbasedtest} and MuTAP~\cite{dakhel2023effectivetestgenerationusing} target focal method generation, using coverage and mutation score as feedback. Unfortunately, due to targeting each method individually, both methods do not scale well to large repositories, becoming costly to use (average code file in TestGenEval contains 58 focal methods, with each method requiring many iterations with the agent). CoverUp~\cite{pizzorno2025coverupcoverageguidedllmbasedtest} and HITS~\cite{wang2024hitshighcoveragellmbasedunit} extend the prior work by adding dependency analysis and error report feedback, but still suffer from cost issues due to their method level approach. \toolname takes a more fluid approach, enabling the LLM to view dependencies as it sees fit, while also allowing it to generate multiple tests at a time.

% \clg{I moved the below from the introduction --- it was atoo long.  Maybe it can just be cut, I'm bringing it down here for you to check to see if any detail should be retained.  Key goal: make us sound novel wrt CoverUp especially (the description in this paragraph makes it sound like we're not, but something in our Slack convo this morning reconvinced me that we are.}
% For instance, CoverUp~\cite{pizzorno2025coverupcoverageguidedllmbasedtest} interleaves dynamic coverage analysis with dialog-based interactions with the LLM by continuously incorporating execution feedback—such as compilation errors, runtime failures, and specific lines or branches lacking coverage—into successive prompts. Similarly, HITS~\cite{wang2024hitshighcoveragellmbasedunit} decomposes complex methods into smaller slices and iteratively prompts an LLM to generate tests for each slice, effectively simplifying analysis scope and increasing branch and line coverage. 

\section{Conclusion}
\label{sec:conclusion}

In this work, we introduced \toolname, an agentic, feedback-driven test generation framework that iteratively refines test suites based on execution results and coverage reports. By leveraging an agentic loop, \toolname effectively balances test readability, coverage, and cost efficiency, outperforming both search-based and LLM-based baselines. Our experiments on TestGenEval demonstrate a significant improvement in pass@1 (84.3\%), code coverage (44.4\%), and mutation score (33.8\%), while maintaining cost-efficiency (\$0.63 per test file).  The produced tests are also syntactically more similar to human-produced tests than prior classic automated approaches. The agentic feedback loop enables dynamic adaptation, refining tests iteratively, to address both coverage gaps and mutation score deficiencies. Our ablation studies further highlight the benefits of starting with zero-shot prompting and iterating on execution feedback, key design decisions. By integrating with OpenHands, we hope to encourage future advancements in test generation research and foster more effective and scalable automated testing solutions.\footnote{Anonymized replication: https://anonymous.4open.science/
r/OpenHands-7E28/}

\section{Data Availability} We release both the agentic version of TestGenEval and the prompt and implementation of \toolname at \url{https://anonymous.4open.science/r/OpenHands-7E28/}. The README detailing how to reproduce all of our experiments is available at \\ \texttt{evaluation/benchmarks/testgeneval/README.md}.

\balance

\bibliographystyle{ACM-Reference-Format}
\bibliography{bibliography_formatted}

\balance

\end{document}